\documentclass[11pt]{article}
\usepackage[margin=1in]{geometry}
\usepackage{enumerate}
\usepackage{tikz}
\usepackage{amsmath}
\usepackage{amsfonts}
\usepackage{float}
\usetikzlibrary{shapes.misc, positioning}
\usetikzlibrary{arrows.meta, positioning, quotes}
\usetikzlibrary{patterns}
\usepackage{hyperref}

\usepackage{cite}

\usetikzlibrary{automata,positioning}

\begin{document}

\title{Fully Homomorphic Image Processing}

\author{William Fu, Raymond Lin, Daniel Inge \\ Harvard University \\ \{wfu, rlin, dinge\}@college.harvard.edu}
\date{May 2, 2018}

\maketitle

\section{Introduction}
Fully homomorphic encryption \cite{Gentry:FHE}, \cite{Gentry:FHE2} has allowed devices to outsource computation to third parties while preserving the secrecy of the data being computed on. Many images contain sensitive information and are commonly sent to cloud services to encode images for different devices. We implement image processing homomorphically that ensures secrecy of the image while also providing reasonable overhead. The paper is organized as follows. First, we will present some previous related work. Next, we will present the fully homomorphic encryption scheme we use. Then, we will introduce our schemes for JPEG encoding and decoding, as well as schemes for bilinear and bicubic image resizing. We then present some data and analysis of our homomorphic schemes. Furthermore, we outline several issues with the homomorphic evaluation of proprietary algorithms, and how a third party can gain information on the algorithm through noise, followed by some concluding remarks. Source code is available at: \url{https://github.com/wfus/Fully-Homomorphic-Image-Processing}. 

\section{Related Work}

Ever since fully homomorphic encryption \cite{Gentry:FHE}, \cite{Gentry:FHE2} was introduced, researchers have worked to practically apply it in a variety of applications. In recent years, such work has included homomorphic biomedical analysis \cite{FHE:Bio} and evaluation of encryption circuits \cite{FHE:AES}, neural networks \cite{FHE:CryptoNets}, and discrete Fourier transforms \cite{FHE:DFT}.

Fully homomorphic encryption's use to perform secure computations on sensitive medical, financial, and other types of data is particularly important in the context of cloud computing, where a party with sensitive information would like to outsource a computation to the cloud while still retaining privacy. Fully homormophic computation has been used to calculate sensitive statistics for biomedical analysis \cite{FHE:Bio}. More complicated algorithms such as edit distance has also been run homomorphically for use in calculating genetic distances and for the use in genomic sequence analysis \cite{FHE:EditDistance}.

\begin{figure*}[h]
\begin{center}
\begin{tikzpicture}

\draw[thick] (0, 0) rectangle node[text width=1cm, align=center] {Raw \\ Image} ++(2, 2);
\draw[thick] (3, 0) rectangle node[text width=2cm, align=center] {Encrypted \\ Image} ++(2, 2);
\draw[thick, color={rgb:red,1;green,2;blue,5}] (2.5, -1) rectangle node[text width=1cm, align=center] {} ++(9, 5);
\draw[thick] (6, 1) rectangle  node[text width=1cm, align=center] {Size \\ 1} ++(1.5, 1.5);
\draw[thick] (6.3, -0.5) rectangle node[text width=1cm, align=center] {Size \\ 2} ++(1, 1);
\draw[thick] (8, 1) rectangle  node[text width=4cm, align=center] {FHE Neural Net} ++(3, 1.5);
\draw[thick] (8, -0.5)  rectangle node[text width=4cm, align=center] {FHE Neural Net} ++(3, 1);
\draw[thick] (12, 1.25)  rectangle node[text width=4cm, align=center] {$y_1$} ++(1, 1);
\draw[thick] (12, -0.5)  rectangle node[text width=4cm, align=center] {$y_2$} ++(1, 1);
     
\draw[thick,->] (2,1) -- ++(1, 0);     
\draw[thick,->] (7.5,1.75) -- ++(0.5, 0);     
\draw[thick,->] (7.3,0) -- ++(0.7, 0);     
\draw[thick,->] (11,1.75) -- ++(1, 0);     
\draw[thick,->] (11,0) -- ++(1, 0);     
\draw[dashed] (5, 2) -- (6, 2.5); 
\draw[dashed] (5, 0) -- (6, 1); 
\draw[dashed] (5, 2) -- (6.3, 0.5); 
\draw[dashed] (5, 0) -- (6.3, -0.5); 

\draw[color={rgb:red,1;green,2;blue,5}] (7, 3.3) node[align=center] {Homomorphically Evaluated};
     
\end{tikzpicture}

\title{Image Resizing Potential Workflow}
\caption{A potential application for image resizing. Many neural nets accept a statically sized input. Resizing an image homomorphically would allow a cloud service to run several prediction algorithms on the same image while ensuring that the user will only have to encrypt the image once, since encryption is an expensive operation for end devices.}
\end{center}
\end{figure*}
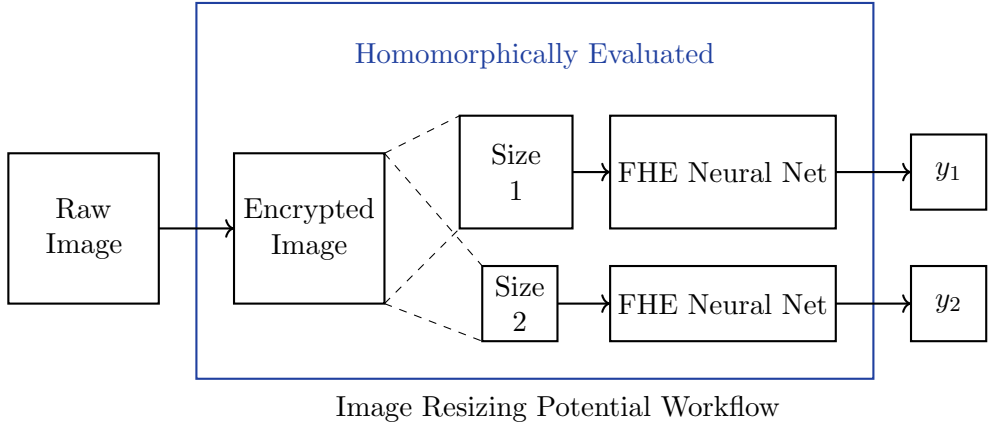

Unfortunately, most homomorphic computation models suffer from several issues. First, a client that wants to outsource computation on sensitive data will usually not have much computation power themselves. However, the encryption step takes a nontrivial amount of computational resources. As in the example of neural networks  \cite{FHE:CryptoNets}, a client can send a sensitive image through neural network homomorphically to get a prediction. In order to outsource a prediction for an image, the client will have to encrypt the entire uncompressed image and send through the cloud. Furthermore, many neural networks take in a fixed sized image for prediction, or sample different areas of the image. Therefore, we implement homomorphic image resizing and homomorphic image decompression. Decompressing images homormorphically (for example from the JPEG) format would allow clients to have to homomorphically encrypt significantly less data, since the client most likely does not have abundant computational resources.

\section{Homomorphic Encryption Scheme}

In order to implement homomorphic encryption and evaluation, we use Microsoft's SEAL library \cite{SEAL:Main}. We will discuss briefly how numbers are encoded in SEAL, as well as the construction and properties of the homomorphic encryption scheme SEAL uses.
\subsection{Encoding}

In SEAL, numbers are encoded as polynomials in the ring $R_t = \mathbb{Z}_t / (x^n + 1)$ where $t \in \mathbb{N}$ is called the plaintext modulus and $n$ is a power of $2$. 

As we perform operations on real numbers rather than just integers in the course of our implementations, we convert numbers to polynomials using the fractional encoding scheme in SEAL. This involves choosing some base $B$, computing the representation of that number in that base, and encoding that representation into a polynomial. To encode a number $$y = \text{sign}(y) \left(\ldots+ b_3 B^{-3} + b_2 B^{-2} + b_1 B^{-1} + a_0 + a_1 B + a_2 B^2 + \ldots\right)$$
we would use the polynomial $$\text{sign}(y) \left(\ldots - b_3 x^{n-3} - b_2 x^{n-2} - b_1 B^{n-1} + a_0 + a_1 x + a_2 x^2 + \ldots\right)$$ 
Note that some numbers might not have a finite representation in a given base, and we have a limited number of coefficients for in our polynomial. As such SEAL provides parameters $n_i$ and $n_f$ that set the maximum number of coefficients to be used for the integer ($a_i$ coefficients) and fractional ($b_i$ coefficients) parts of the representation, respectively.

\subsection{Encryption}

The homomorphic encryption scheme used in SEAL is due to Fan and Vercauteren \cite{FHE:FV} with slight modification to the decryption, addition, and multiplication operations \cite{SEAL:Main}. Plaintexts are polynomials in $R_t$ as defined in the above encoding section, and ciphertexts are arrays of polynomials in $R_q = \mathbb{Z}_q/(x^n +1)$ for $q \in \mathbb{N}$ is called the ciphertext modulus and $n$ is as defined earlier.

Let us define $\lambda$ to be a security parameter, $R_2 = \mathbb{Z}_2 / (x^n+1)$, distribution $\chi$ defined based on the discrete Gaussian, $[\cdot]_q$ to be reduction of polynomial coefficients modulo $q$, $\lfloor \cdot \rceil$ to be integer rounding, $\lfloor \cdot \rfloor$ to be integer floor, and $\lceil \cdot \rceil$ to be integer ceiling. Key generation is defined as
$$\begin{aligned}\mathsf{KeyGen}(\lambda): & s \overset{R}{\leftarrow} R_2, a \overset{R}{\leftarrow} R_q, e \overset{R}{\leftarrow} \chi \\
& (sk, pk) \leftarrow (s, ([-(as + e)]_q, a))
\end{aligned}$$
Encryption and decryption are defined as
$$\begin{aligned}\mathsf{Encrypt}(pk, m): & u \overset{R}{\leftarrow} R_2, e_0, e_1 \overset{R}{\leftarrow} \chi (p_0, p_1) \leftarrow pk \\
& ct \leftarrow \left(\left[\left\lfloor \frac{q}{t} \right\rfloor m + p_0 u + e_0 \right]_q, [p_1 u + e_1]_q \right) \\
\mathsf{Decrypt}(sk, ct): & s \leftarrow sk, (c_0, \ldots, c_k) \leftarrow ct \\
& m \leftarrow \left[\left\lfloor \frac{t}{q}\left[\sum_{i=0}^k c_i s^i\right]_q \right\rceil \right]_t
\end{aligned}$$
Finally, homomorphic addition and multiplication are performed by doing
$$\begin{aligned}\mathsf{Add}(ct_0, ct_1): & (c_0,\ldots,c_j) \leftarrow ct_0, (d_0,\ldots,d_k) \leftarrow ct_1 \\
&ct_2 \leftarrow (c_0 + d_0, \ldots, c_{\max(j, k)} + d_{\max(j, k)}) \\
\mathsf{Multiply}(ct_0, ct_1): & (c_0,\ldots,c_j) \leftarrow ct_0, (d_0,\ldots,d_k) \leftarrow ct_1 \\
&ct_2 \leftarrow \left(\left[\left\lfloor\frac{t}{q} \sum_{r+s=0} c_r d_s \right\rceil \right]_q, \ldots, \left[\left\lfloor\frac{t}{q} \sum_{r+s=j + k} c_r d_s \right\rceil \right]_q\right)
\end{aligned} $$

We will now briefly state a few basic properties of the encryption parameters and evaluation operations. First off, the noise budget cost of addition or subtraction operations is very low compared to that of a multiplication operation. Increasing the ciphertext modulus $q$ increases the initial noise budge of a freshly encrypted ciphertext as well as decreases the security. Increasing the power of the polynomial modulus $n$ increases security while decreasing speed. In addition, increasing the plaintext modulus $t$ decreases the noise budget and increases the cost of multiplicative operations, while it allows for computations on larger numbers or with more precision.

As multiplication increases the size of the ciphertext by increasing the number of polynomials in the ciphertext array, such operations result in higher computation time needed for further operations and high consumption of the noise budget. To combat this, the FV scheme also includes an operation called relinearization that reduces the ciphertext size. However, as the multiplicative depth of the circuits we use are not particularly high, and relinearization itself has a computation cost (as well as a noise budget cost comparable to that of addition or subtraction), we do not make use of it in our implementation.

\section{Homomorphic Image Processing}

\subsection{Image Resizing}
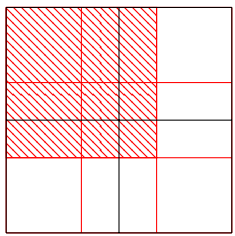
\begin{figure}[h]
\begin{center}
\begin{tikzpicture}
\draw[step=1cm,color=red] (0,0) grid (3,3);
\draw[pattern=north west lines, pattern color=red, draw=red] (0,1) rectangle (2,3);
\draw[step=1.5cm,color=black] (0,0) grid (3,3);
\end{tikzpicture}
\end{center}
\vspace{-5mm}
\caption{Finding nearby pixels in the original image for a pixel in the new image. The red lines are the pixels of the original image, the black lines are the pixels of the new image, and the pixels in the original picture that are dotted are the $2 \times 2$ square of pixels in the old image that would be ``nearby" to the pixel in the upper left corner of the resized image.}
\label{fig:interpolation}
\end{figure}

When we resize an image, we must have a way of defining each of the pixels in the new image based on the pixels of the old image. A common way of doing this is using the pixels in the original image that would be around a pixel in the new image to perform some sort of interpolation. This is illustrated in Figure \ref{fig:interpolation}. 
\vspace{5mm}

We implement two types of interpolation commonly used for image rescaling: bilinear and bicubic interpolation.

\begin{figure}[h]
\begin{center}
\includegraphics[scale=0.45]{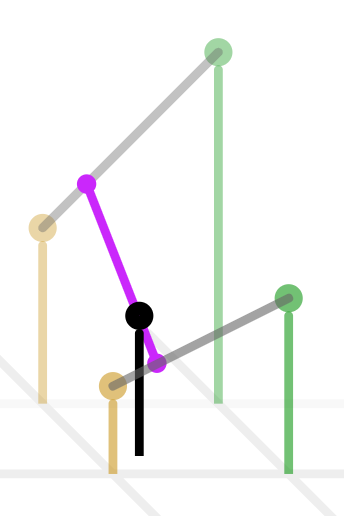}
\end{center}
\vspace{-5mm}
\caption{Bilinear interpolation. The green and yellow points are points where we know the value of the function, and the black point is the one which we wish to find, and the heights of the points above the plane represent the function value (image obtained from \cite{Resize:Comparison}).}
\label{fig:linear}
\end{figure}

Bilinear interpolation involves performing linear interpolation in two dimensions. Linear interpolation uses a linear approximation to guess the for the value of a function at a point between points with known function value. The formula for linearly interpolating the value of function $f$ at $x$ between $x_0$ and $x_1$, if $t = \frac{x - x_0}{x_1-x_0}$, is 
\begin{equation}f(x) = \begin{pmatrix}1 & t \\ \end{pmatrix} \begin{pmatrix} 1 & 0 \\ -1 & 1 \\ \end{pmatrix} \begin{pmatrix} f(x_0) \\ f(x_1) \end{pmatrix}
\end{equation}
To extend to bilinear interpolation, we first interpolate values in one dimension, and then in the other, as illustrated Figure \ref{fig:linear}. 
\vspace{5mm}

Each linear interpolation requires two points, so we will need the $2 \times 2$ square of pixels around the center of the new pixel to perform our interpolation.

\begin{figure}[h]
\begin{center}
\includegraphics[scale=0.4]{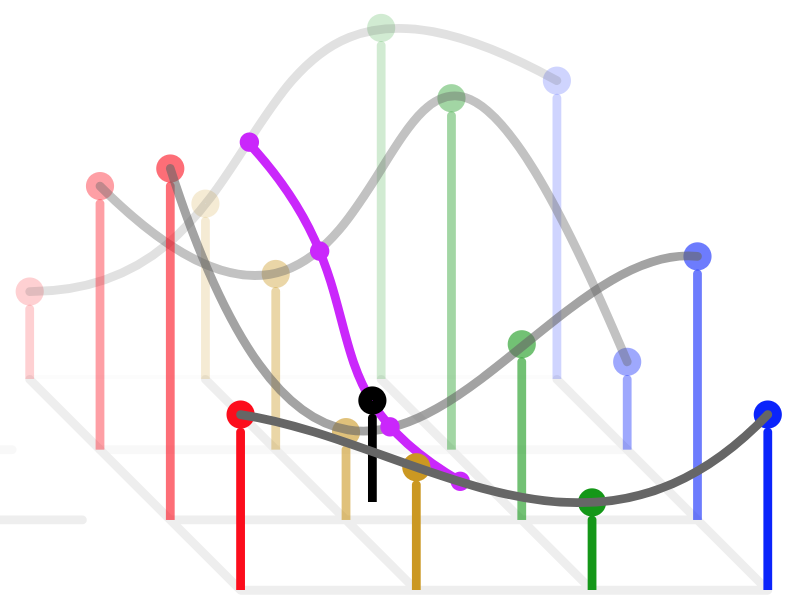}
\end{center}
\vspace{-5mm}
\caption{Bicubic interpolation. The red, yellow, green and blue points are points where we know the value of the function, the black point is the one which we wish to find, and the heights of the points above the plane represent the function value (image obtained from \cite{Resize:Comparison}).}
\label{fig:cubic}
\end{figure}

Bicubic interpolation, on the other hand, performs cubic interpolation in two dimensions. Linear interpolation uses a cubic approximation to guess the for the value of a function at a point between points with known function value. The formula for interpolating the value of function $f$ at $x$ between $x_0$ and $x_1$, if $t = \frac{x - x_0}{x_1-x_0}$, $x_{-1} = 2x_0 - x_1-x_0$, $x_2 = 2x_1 - x_0  $ is \begin{equation}f(x) = \frac{1}{2}\begin{pmatrix}1  & t & t^2 & t^3\\ \end{pmatrix}
\begin{pmatrix}
0 & 2 & 0 & 0\\
-1 & 0 & 1 & 0 \\
2 & -5 & 4 & -1 \\
-1 & 3 & -3 &1 \\
\end{pmatrix} \begin{pmatrix} f(x_{-1}) \\ f(x_0) \\ f(x_1) \\ f(x_2) \end{pmatrix}
\end{equation}
To do bicubic interpolation, we first interpolate values in one dimension, and then in the other, as illustrated in Figure \ref{fig:cubic}.  

\vspace{5mm}
Since each cubic interpolation requires four points, we will need the $4 \times 4$ square of pixels around the center of the new pixel to perform our interpolation.

Since bilinear and bicubic interpolation are built up directly from linear and cubic interpolation, respectively, our homomorphic implementation simply involved writing functions to homomorphically evaluate those formulas.

\subsection{Image Compression}

We implement JPEG encoding, which is usually done by performing a YCbCr color transform, a discrete cosine transform (DCT), quantization, and lossless entropy encoding on each $8 \times 8$ block in the original image. We will briefly describe each of these steps here.

\begin{figure*}[h]
\begin{center}
\begin{tikzpicture}
[
    node distance=5mm and 4mm,
    box/.style = {draw, minimum height=15mm, inner xsep=3mm, align=center},
    sy+/.style = {yshift= 2mm},
    sy-/.style = {yshift=-2mm},
    every edge quotes/.style = {align=center},
]
     \node (n1) [box]             {Raw\\Image};
     \node (n2) [box,right=of n1] {Color \\ Transform \\ (RGB-YUV)};
     \node (n3) [box,right=of n2] {DCT};
     \node (n4) [box,right=of n3] {Quantization};
     \node (n5) [box,right=of n4] {Entropy\\Encoding};
     \node (n6) [box,right=of n5] {JPEG\\Headers};
     
     \draw[-Triangle] (n1.east) to (n2.west);
     \draw[-Triangle] (n2.east) to (n3.west);
     \draw[-Triangle] (n3.east) to (n4.west);
     \draw[-Triangle] (n4.east) to (n5.west);
     \draw[-Triangle] (n5.east) to (n6.west);
     \draw[thick,color={rgb:red,1;green,2;blue,5}] (1, -1.3) rectangle (9.2, 2);
     \draw[color={rgb:red,1;green,2;blue,5}] (5.2, 1.1) node[above] {Homomorphically Evaluated};
     
\end{tikzpicture}

\title{JPEG Standard Image Compression}
\caption{The workflow for JPEG compression using our homomorphic scheme.}
\end{center}
\end{figure*}
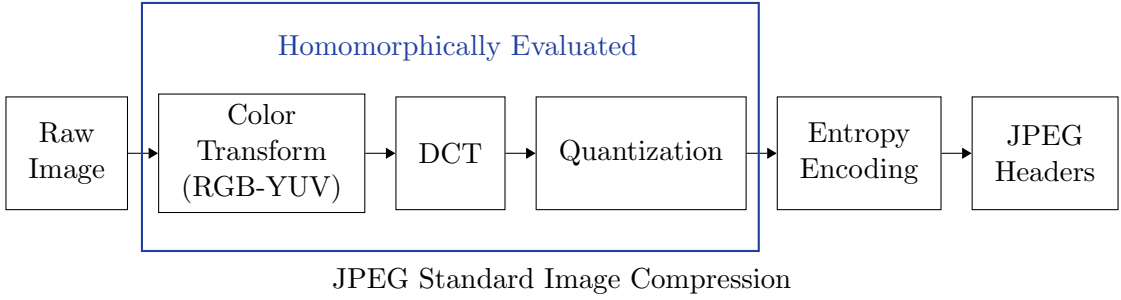

The Y'CbCr color transform involves taking our representation of color for each pixel from the standard RGB space to the Y'CbCr space, where Y' is a luma value and Cb, Cr are chroma values. We move to this color space since human visual perception is more sensitive to brightness than to color, so this representation is more compressible than is RGB color specification. The transformation between the color spaces is given by $$\begin{pmatrix}Y' \\ Cb \\ Cr \end{pmatrix} = 
\begin{pmatrix} 0.299 & 0.587 & 0.114 \\
-0.168736 &- 0.331264 & 0.5 \\
0.5 & -0.418688 & -0.081312\end{pmatrix}
\begin{pmatrix}R \\ G \\ B \end{pmatrix}$$

The DCT is a frequency domain representation of the image. If $A$ is a $8 \times 8$ block for one channel of the color space and $B$ is the resultant $DCT$ matrix, we have that if $\alpha(x)$ is $\frac{1}{\sqrt{2}}$ if $x=0$ and $1$ otherwise, $$B_{uv} = \frac{1}{4} \alpha(u) \alpha(v) \sum_{x=0}^7 \sum_{y=0}^7 A_{xy} \cos\left(\frac{2x+1}{16}\pi u \right) \cos\left(\frac{2y+1}{16}\pi v \right)$$

The quantization process is where the lossy part of JPEG compression occurs. Since DCT coefficients for high frequencies are often small values, and the disposal of such frequencies would lead to little perceptible difference, we divide the DCT coefficients by elements of a quantization matrix $Q$ and round the result. If $C$ is the resultant matrix after quantization of $B$ using quantization matrix $Q$, we have that $$C_{ij} = \left\lfloor \frac{B_{ij}}{Q_{ij}}\right\rceil$$

Finally, the lossless entropy encoding involves ``zig-zagging" through the quantized DCT coefficient matrix and performing run length encoding on that sequence, followed by Huffman encoding on the result of that.

It goes against CPA security to implement a reasonable entropy encoding and send it back to the client with less data (it would be possible to do the encoding, but the server would have to send back padded zeros anyway with the same size). As the entropy encoding takes less time than the color transform, DCT, and quantization steps, we implement homomorphic evaluation for all of the steps except for the entropy encoding, which the client performs after receiving the result from the other steps from the server.

\subsection{Image Decompression}

We implemented simple run length decoding, since most lossy image formats use run length encoding after a DCT as the main compression technique. We tested with simple run length decoding (without the Huffman coding using in the JPEG file format) consisting of pairs $(a_i, b_i)$ where the pair corresponds with $a_i$ being repeated $b_i$ times. An example sequence is 
\[ [(A, 8), (B, 3), (C, 5)] \mapsto [ AAAAAAAABBBCCCCC ] \]

Since the H.264 and MPEG video encoding standard uses run length encodings that encode lengths of $16$ ($ 4 \times 4$ blocks), $64$ ($ 8 \times 8$ blocks), and $256$ ($ 16 \times 16$ blocks)and JPEG uses run length encodings to encode lengths of $64$ ($8 \times 8$ blocks), we implemented our homomorphic run length decoding for output lengths of $16$ and $64$. 

There are methods of compiling a function into boolean circuits \cite{FHE:Circuit} to take advantage of conditionals. However, converting the run length decoding into a boolean circuit would not allow the resulting image to be used in a homomorphic neural network, so we investigated ways to approximate run length decoding without converting into a boolean circuit. Unfortunately, the homomorphic encryption scheme does not allow for conditional expressions or division by ciphertexts. Therefore, we use the following method of approximating the run length encoding. Consider an example with a fixed output length of $16$ as in the case for video encodings. 
\[ [(20, 8), (30, 3), (15, 5)] \mapsto [ 20, 20, 20, 20, 20, 20, 20, 20, 30, 30, 30, 15, 15, 15, 15, 15 ] \]
Since the output array is a fixed length, we can image that is it a continuous function $f$ with domain from $0$ to $16$, and split the function apart into $f_1, f_2, f_3$ corresponding to each pair, which is depicted as follows. 

\begin{center}
\[ \hspace{-1cm}
\includegraphics[scale=0.35]{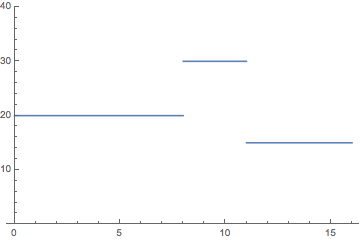}
\hspace{5mm}
\includegraphics[scale=0.35]{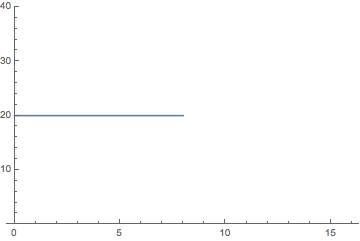}
\hspace{5mm}
\includegraphics[scale=0.35]{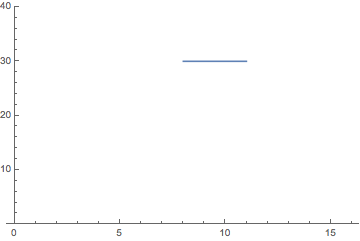}
\hspace{5mm}
\includegraphics[scale=0.35]{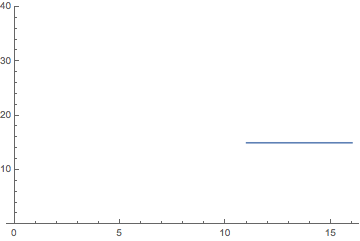}
\]
\[ \hspace{-1cm}
f(x) = 
\begin{cases} 
    0 & x < 0 \\ 
    20 & 0 < x < 8 \\ 
    30 & 8 < x < 11 \\ 
    15 & 11 < x < 16 \\ 
    0 & x > 16 
\end{cases} 
~ ~ ~ ~ ~
f_1(x) = 
\begin{cases} 
    0 & x < 0 \\ 
    20 & 0 < x < 8 \\ 
    0 & x > 8 
\end{cases} 
~ ~ ~ ~ ~
f_2(x) = 
\begin{cases} 
    0 & x < 8 \\ 
    30 &  8 < x < 11 \\ 
    0 & x > 11 
\end{cases} 
~ ~ ~ ~ ~
f_3(x) = 
\begin{cases} 
    0 & x < 0 \\ 
    20 & 11 < x < 16 \\ 
    0 & x > 16 
\end{cases} 
\]
\end{center}

\begin{figure}[h]
    \begin{center}
    \hspace{-1.8cm}
    \begin{minipage}[c]{0.35\textwidth}
    \begin{center}
    \includegraphics[scale=0.33]{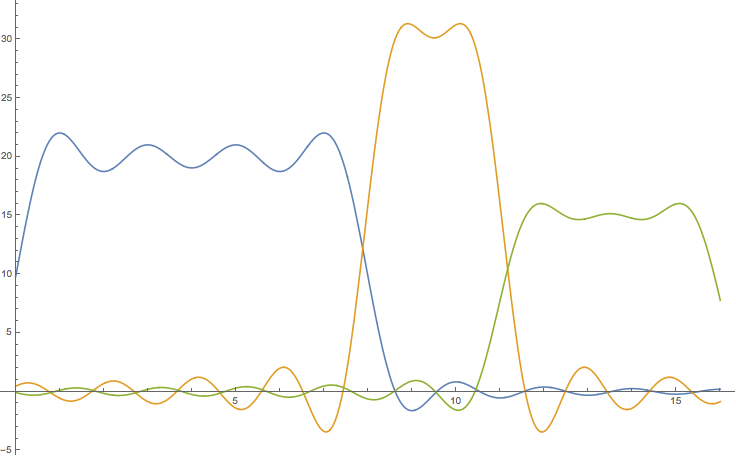}
    \end{center}
    \end{minipage}
    \hspace{3cm}
    \begin{minipage}[c]{0.35\textwidth}
    \begin{center}
    \includegraphics[scale=0.33]{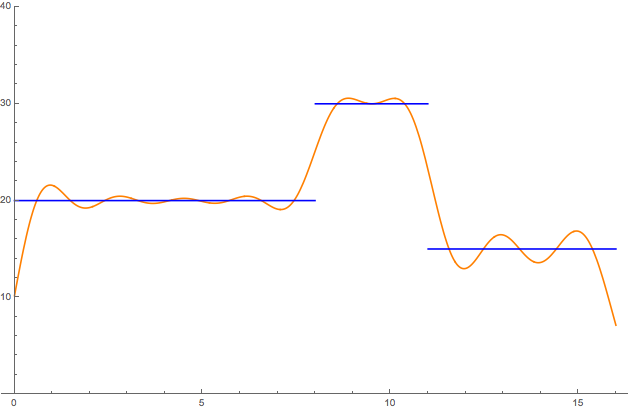}
    \end{center}
    \end{minipage}
    \end{center}
    \caption{The separate Fourier series for the the functions $f_1$, $f_2$, $f_3$ using our example $[(20, 8), (30, 3), (15, 5)]$. The Fourier series is centered at $0$ with a period of $64$, approximated to the $16^\text{th}$ term. Then, the series is shifted on the $x$ axis to get the peaks to the correct position. The sum $f = f_1 + f_2 + f_3$ is plotted on the right and is compared to the original step functions it approximates.}
    \label{fig:fourierseries}
\end{figure}

However, it is difficult to recreate these step functions without using conditionals. It is also difficult to use a sigmoid type function to approximate these step functions, because sigmoid involves ciphertext division. We approximated the step functions using the Fourier series, since all of the coefficients do not require ciphertext division and since the sine and cosine functions can be Taylor expanded to use only ciphertext multiplication. However, since we have to take the Taylor expansion of sine and cosine, we have to try to minimize the domain of the inputs into sine and cosine. Therefore, we approximate using the Fourier series with a periodicity of $64$ and our step functions. The Fourier series for this example is explicitly plotted in Figure \ref{fig:fourierseries}.

To evaluate our actual output array and the sixteen decoded values, we will sum up the Fourier series of $f_1$, $f_2$, and $f_3$ to find $f$, which is shown in Figure \ref{fig:fourierseries}. Then, our output array will just be equal to the function at $0, \ldots, 15$, and the decoded sequence will be 
\[ \mathsf{Decoded Sequence} =  [f(0),\ldots, f(15)]\]

Note that it is possible to do run length decoding homomorphically because the Fourier series of the step function defined as $1$ on $(-b, b)$ and $0$ elsewhere periodic from $0$ to $64$ is just
\[ H(b, o, x) = \frac{b}{64} + \sum_{k = 0}^{\mathsf{deg}} \frac{2}{k\pi} \sin\left( \frac{k b \pi}{64} \right) \cos \left (\frac{k (x - o) \pi }{64} \right) 
\]
Then, given a run length encoding given by $[(a_1, b_1), \dots, (a_t, b_t)]$, the series representation of our step functions will be shifted versions of $H$. Then, we will have 
\[ f(x) = H_1(x) + \cdots + H_t(x) ~ ~ ~ ~  ~ ~ ~ ~ ~ ~ ~ H_i(x) = a_i  \cdot H\left(\frac{b_i}{2} + \delta - \frac{1}{2}, \frac{b_i}{2}+ \sum_{k = 1}^{i-1} b_k, x   \right)\]
where $\delta$ is a parameter that affects the width of each step of our approximated step function.

We use the Taylor expansion of $\sin$ and $\cos$ to the $10^\text{th}$ order term to get a larger domain to work with, since it is impossible to take a number modulo $2\pi$ homomorphically. Since all of these steps use the ciphertexts $a_i, b_i$ in the numerator and using only addition and multiplication, it works homomorphically.

We note when applied to the JPEG decoding process, the noise resulting from our approximated step functions should not prove to be a significant problem. As JPEG is a lossy compression scheme, we see that the noise, when not excessive, should not have a significant impact on the quality of the resultant image.

\section{Evaluation}

\subsection{Timing}

For benchmarking purposes, we ran all the algorithms on a computer with a 12 core Intel i7-8700K with a clock rate of $3.70$ GHz with 16 GB of DDR4 RAM. As $n$ the power of the polynomial modulus has the most significant impact of the speed \cite{SEAL:Main}, we found the average runtime of the encryption and decryption operations in the homomorphic encryption scheme as well as the most expensive operations we implemented for image resizing and compression as a function of $n$. This averaging was done over multiple runs of each of the functions, as well as variations in plaintext modulus. The resultant timings for the various operations can be found in Figure \ref{fig:timing}.

\vspace{5mm}
\begin{figure}[h]
\begin{center}
\begin{tabular}{| l | c | c | c | c |}
\hline
& $n=2048$ & $n=4096$ & $n=8192$ & $n=16384$ \\
\hline
Encryption & 0.851 & 1.782 & 3.978 & 11.223 \\
\hline
Decryption & 0.079 & 0.268 & 1.003 & 4.233 \\
\hline
Linear Interpolation & 3.057 & 10.399 & 39.418 & 170.172 \\
\hline
Cubic Interpolation & 9.078 & 31.664 & 122.357 & 526.084 \\
\hline
DCT & 55.701 & 199.171 & 762.647 & 3093.471 \\
\hline
\end{tabular}
\caption{The average runtime in milliseconds for encryption and decryption in the homomorphic scheme, as well as our various homomorphically evaluated functions, as a function of $n$, the power of the polynomial modulus.}
    \label{fig:timing}
\end{center}
\end{figure}

\subsection{Image Resizing}

\begin{figure}[h]
    \begin{center}
    \begin{tabular}{c c c}
         \includegraphics[scale=5]{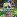}
         &  
         \includegraphics[scale=5]{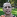}
         &
         \includegraphics[scale=5]{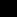}
         \\
          t=11 & t=1009 & t=100003
    \end{tabular}
    \end{center}
    \vspace{-5mm}
    \caption{Bicubic image resizing on a picture of Professor Boaz Barak to a $17\times 17$ image with polynomial modulus $x^{4096} + 1$ and plaintext modulus $t$.}
    \label{fig:bicubic}
\end{figure}

Varying encryption parameters had similar effects on both bilinear and bicubic resizing. Since bicubic resizing requires greater multiplicative depth, and the images produced with bilinear interpolation were quite similar to the ones from bicubic inteprolation, we will show the results we obtained during benchmarking for bicubic resizing. We resized a $48 \times 48$ image to a destination size of $17 \times 17$. Some results while varying the plaintext modulus $t$ is shown in Figure \ref{fig:bicubic}. 

\vspace{5mm}

Since we are working in a modular system, it is possible for operations to get corrupted due to overflow when numbers exceed the modulus size. We can see that setting a smaller modulus size of $t = 11$ produces visible overflowed pixels. However, using too large of a plaintext modulus space decreases the amount of operations we can do, as this results in a decrease in the initial noise budget of a ciphertext, as well an increase noise cost of performing multiplicative operations on ciphertexts, thereby resulting in completely indecipherable ciphertexts. The result of such excess of the noise budget can be seen in the black image resulting with $t = 100003$.

\subsection{Image Compression}

\begin{figure}[h]
    \begin{center}
    \begin{tabular}{c c c c c }
         \includegraphics[scale=1.6 ]{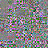}
         &  
         \includegraphics[scale=1.6]{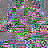}
         &
         \includegraphics[scale=1.6]{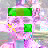}
         &
         \includegraphics[scale=1.6]{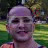}
         &
         \includegraphics[scale=1.6]{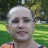}
         \\
         $t = 11$ & $t = 31$ &$ t = 101$ & $t = 307$ & $t=1009$
    \end{tabular}
    \end{center}
    \vspace{-5mm}
    \caption{JPEG encoding on a picture of Professor Boaz Barak. All of the pictures were computed with polynomial modulus $x^{2048} + 1$ and plaintext modulus $t$.}
    \label{fig:encoding}
\end{figure}

We observed similar effects with overflow as in the previous section about image resizing. However, in this case, we can see that each overflow will only have a corrupting effect on the localized $8 \times 8$ where it is located. We can see the corrupted high frequencies in Figure \ref{fig:encoding}. Note that the high frequency components of the DCT do not usually overflow, since we can see the outline of Professor Barak's face in the $t = 11$ case even as all of the chroma elements have been corrupted. 

\vspace{5mm}

\subsection{Image Decompression}

\begin{figure}[h]
    \begin{center}
    \begin{tabular}{c c c c c c}
         \includegraphics[scale=0.45]{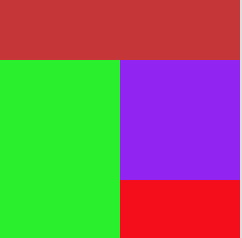}
         &  
         \includegraphics[scale=0.45]{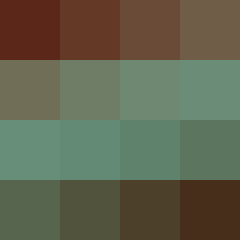}
         &
         \includegraphics[scale=0.45]{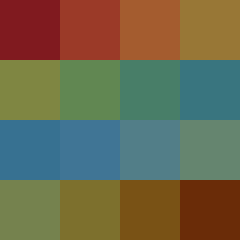}
         &
         \includegraphics[scale=0.45]{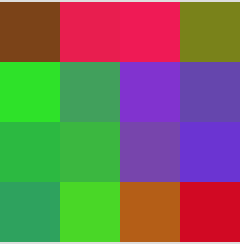}
         &
         \includegraphics[scale=0.45]{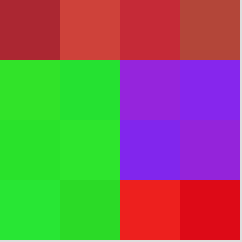}
         &
         \includegraphics[scale=0.45]{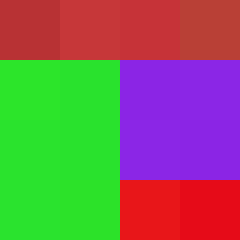}
         \\
         Original Image &  
         8 Terms &  
         16 Terms &
         32 Terms &
         48 Terms &
         64 Terms
    \end{tabular}
    \end{center}
    \vspace{-5mm}
    \caption{The effect of our image decompression approximation given different amounts of terms used in the Fourier series expansion. We used the decompression assuming the run length encoding was used for RGB values, which causes the smoothing effect. Normally, the run length encoding in actual the actual JPEG and video formats are done after DCT and in the YUV space.}
    \label{fig:decode}
\end{figure}

When running the image decompression algorithm, several parameters needed to be tuned for correctness. Since we use a Fourier series to approximate the step functions, the approximation must go from $0$ to $1$ continuously. Therefore, using the Fourier series will cause values near the boundaries of the run length segments to bleed over, generating a smoothing effect, which becomes more pronounced as less Fourier series terms are used. Some resulting images with different numbers of terms in our Fourier series are presented in Figure \ref{fig:decode}.

\vspace{5mm}

We observe that as we increase the number of terms, the Fourier approximation gets better and better, and hence the amount of blending that occurs decreases. Once we get to 64 terms, the resultant picture is already extremely close to the original.

\section{Functional Privacy}

Although homomorphic encryption schemes give some security guarantees for ciphertexts, they do not necessarily provide obfuscation of the function performed on the server. For instance, with careful choice of initial noise budget in a ciphertext submitted for computation on a proprietary neural network, a client might be able to exfiltrate information on the number of layers in the neural network based on noise corruption in the result obtained from the server, since multiplicative operations take a heavy toll on noise budget. 

Given that some of our work's primary applications lie in application of homomorphic image processing tools for server-side computation on client data, we will briefly mention some recently proposed resolutions to this problem. Gentry \cite{Gentry:FHE} suggests an approach, where when we receive an input ciphertext, we add it to a ciphertext with value $0$, and noise significantly larger (in fact, super-polynomially larger) than the input ciphertext, to the input. This ``noise flooding" of the input ciphertext effectively destroys the noise information originally present in the input ciphertext. Ducas and Stehle \cite{Attack:Soak} suggest another approach to this issue that does not have the noise tolerance requirements of the Gentry scheme. Their scheme, which they call ``soak-spin-repeat", essentially involves repeated re-encryption with bootstrapping with injections of noise in between re-encryptions in a way similar to the Gentry scheme, except with much smaller noise than is required in that scheme.

\section{Conclusion}

In this report we constructed and benchmarked image resizing, encoding, and decoding using a fully homomorphic scheme. We implemented these using Microsoft's SEAL library assuming a client lacking computational resources wishes to outsource their function to a cloud server. We showed that homomorphic image resizing can feasibly be done on standard size images, even with the limited resources of a standard desktop computer. Furthermore, we have implemented a homomorphic run length decoding algorithm that can be used to implement JPEG and other image decoding algorithms. Currently, in many implementations of neural nets or biometric applications, raw images are encrypted and sent to the server, which consumes a lot of computational power on the client side. Being able to send a compressed image and have it be rescaled and decompressed homomorphically on the server side will significantly decrease the computing power needed on the client side to utilize fully homomorphic encryption schemes, allowing weaker edge devices to securely outsource computation for sensitive data. 

\section{Acknowledgements}

This project in homomorphic encryption was done for CS227 taught by Professor Boaz Barak at Harvard University. We would like to thank him for his lectures and notes (\url{https://www.intensecrypto.org/}) for the course! Also, we would like to thank Chi-Ning Chou and Yueqi Sheng for their support in the class.


\newpage
\begin{thebibliography}{1}



\bibitem{Gentry:FHE}
C. Gentry. Fully Homomorphic Encryption Using Ideal Lattices. In \textit{STOC}, 2009.

\bibitem{Gentry:FHE2}
C. Gentry, S. Halevi, and N.P. Smart.  Fully Homomorphic Encryption with Polylog Overhead. In \textit{Advances in Cryptology - EUROCRYPT}, 2012.

\bibitem{FHE:Bio}
N. Dowlin, R. Gilad-Bachrach, K. Laine, K. Lauter, M. Naehrig, and J. Wernsing. Manual for Using Homomorphic Encryption for Bioinformatics. In \textit{Proceedings of the IEEE}, 2017.

\bibitem{FHE:AES}
C. Gentry, S. Halevi, and N.P. Smart. Homomorphic Evaluation of the AES Circuit. In \textit{Advances in Cryptology – CRYPTO}, 2012.

\bibitem{FHE:CryptoNets}
N. Dowlin, R. Gilad-Bachrach, K. Laine, K. Lauter, M. Naehrig, and J. Wernsing. CryptoNets: Applying Neural Networks to Encrypted Data with High Throughput and Accuracy. In \textit{Proceedings of the 33rd International Conference on International Conference on Machine Learning}, 2016.

\bibitem{FHE:DFT}
A. Costache, N.P. Smart, and S. Vivek. Faster Homomorphic Evaluation of Discrete Fourier Transforms. In \textit{International Conference on Financial Cryptography and Data Security}, 2017.

\bibitem{FHE:EditDistance}
J. Cheon, M. Kim, and K. Lauter. Homomorphic Computation of Edit Distance. In \textit{International Conference on Financial Cryptography and Data Security}, 2015.





\bibitem{SEAL:Main}
H. Chen, K. Han, Z. Huang, A. Jalali, and K. Laine. Simple Encrypted Arithmetic Library v2.3.0. 2017.

\bibitem{FHE:FV}
J. Fan and F. Vercauteren. Somewhat Practical Fully Homomorphic Encryption. In \textit{IACR Cryptology ePrint Archive}, 2012.

\bibitem{Resize:Comparison}
Cmglee. Comparison of 1D and 2D interpolation. 2016. Available: \url{https://commons.wikimedia.org/wiki/FileComparison\_of\_1D\_and\_2D\_interpolation.svg}

\bibitem{FHE:Circuit}
N. Buscher and S. Katzenbeisser. Compilation for Secure Multi-party Computation. Springer International Publishing, 2017. 

\bibitem{Attack:Soak}
L. Ducas and D. Stehle. Sanitization of FHE Ciphertexts. In \textit{Advances in Cryptology - EUROCRYPT}, 2016.
\end{thebibliography}
\end{document}